# METALLICITY EFFECT ON LMXB FORMATION IN GLOBULAR CLUSTERS

D.-W. Kim[1], G. Fabbiano[1], N. Ivanova[2], T. Fragos[1],
A. Jordán[3,4], G. R. Sivakoff[2], and R. Voss[5]

1. Harvard-Smithsonian Center for Astrophysics, 60 Garden Street, Cambridge, MA 02138
2. Department of Physics, University of Alberta, Edmonton, AB, Canada
3. Departamento de Astronomia y Astrofisica, Pontificia Universidad Catolica de Chile, Santiago, Chile
4. The Milky Way Millennium Nucleus, Av. Vicuña Mackenna 4860, 7820436 Macul, Santiago, Chile
5. Depeartment of Astrophysics/IMAPP, Radboud University Nijmegen, The Netherlands

(Nov. 14, 2012)

## ABSTRACT

We present comprehensive observational results of the metallicity effect on the fraction of globular clusters (GC) that contain low-mass X-ray binaries (LMXB), by utilizing all available data obtained with *Chandra* for LMXBs and *HST* ACS for GCs. Our primary sample consists of old elliptical galaxies selected from the ACS Virgo and Fornax surveys. To improve statistics at both the lowest and highest X-ray luminosity, we also use previously reported results from other galaxies. It is well known that the fraction of GCs hosting LMXBs is considerably higher in red, metal-rich, than in blue, metal-poor GCs. In this paper, we test whether this metallicity effect is X-ray luminosity-dependent, and find that the effect holds uniformly in a wide luminosity range. This result is statistically significant (at $\geq 3\sigma$) in LMXBs with luminosities in the range $L_X = 2 \times 10^{37}$ - $5 \times 10^{38}$ erg s$^{-1}$, where the ratio of GC-LMXB fractions in metal-rich to metal-poor GCs is $R = 3.4 \pm 0.5$. A similar ratio is also found at lower (down to $10^{36}$ erg s$^{-1}$) and higher luminosities (up to the ULX regime), but with less significance (~$2\sigma$ confidence). Because different types of LMXBs dominate in different luminosities, our finding requires a new explanation for the metallicity effect in dynamically formed LMXBs. We confirm that the metallicity effect is not affected by other factors such as stellar age, GC mass, stellar encounter rate, and galacto-centric distance.

1. INTRODUCTION

It is well known that globular clusters (GC) are a major birthplace of low-mass X-ray binaries (LMXB) (Grindlay 1984; Verbunt & van den Heuvel 1995; Bildsten & Deloye 2004), where the efficiency of LMXB formation is at least by a factor of ~100 higher than in the general stellar field ('native' LMXBs). While native LMXBs are expected to be rare in GCs (e.g., Ivanova et al. 2008), LMXBs in GCs can be dynamically formed via tidal capture, binary exchanges and physical collisions (e.g., see Verbunt and Lewin 2006 and reference therein) in the dense stellar environment. It was even suggested that all LMXBs in galaxies were formed in GCs (e.g., Grindlay 1984), but there is also evidence for native field binary formation (e.g., Maccarone, et al. 2003; Irwin 2005; Juett et al. 2005; Kim, E., et al. 2006). In particular, the X-ray luminosity functions of LMXBs in the field and in GCs differ at low luminosities (Kim et al. 2009; Voss et al. 2009; Zhang et al. 2011), pointing to different origins.

*Chandra* observations of early type galaxies revealed that a significant fraction (30-70%) of LMXBs are found in GCs and further showed that LMXBs are preferentially found in red rather than blue GCs (e.g., see Fabbiano 2006 and reference therein). Because in old stellar systems, such as GCs, the color is primarily determined by metallicity (e.g., Brodie and Strader 2006), this trend directly indicates that metal-rich GCs generally host LMXBs more frequently than metal-poor ones, as originally observed for GC-LMXBs in the Galaxy and M31 (e.g., Bellazzini et al. 1995). Observations of several galaxies have shown that the fraction of GCs hosting LMXBs (hereafter GC-LMXB fraction) is larger by a factor of ~3 in metal-rich than in metal-poor GCs (e.g., Kundu et al. 2002; Sarazin et al. 2003; Kim E. et al. 2006; Sivakoff et al., 2007; Paolillo et al. 2011). These results have suggested that the GC metallicity must be closely related to key parameters of binary formation and evolution in GCs, opening a debate on its cause. Suggestions include different stellar sizes and/or IMFs (Bellazzini et al 1995), irradiation-induced stellar winds (Maccarone et al. 2004), and metallicity-dependent magnetic breaking (Ivanova 2006). However, none of these possibilities has been observationally supported. For example, the excess obscuration in metal-poor GCs predicted by the stellar wind model was not observed (e.g., Kim, E. et al. 2006). While magnetic breaking could explain the metallicity effect for main-sequence donors, this type of LMXBs is considerably fainter ($L_X < 2 \times 10^{37}$ erg s$^{-1}$) than those observed in typical elliptical galaxies (Ivanova 2006; Revnivtsev et al. 2011).

A key observational test is determining whether the metallicity effect is luminosity-dependent, because different types of LMXBs (with different compact sources and with different donors) dominate in different luminosity ranges (e.g., Fragos et al. 2008). Therefore the luminosity range where the metallicity effect does or does not hold provides an important clue to identify dominant types of LMXBs and to understand the major LMXB formation mechanisms. In this paper, we report the measurement of GC-LMXB fractions in red and blue GCs at different X-ray luminosities, with a large sample of GC-LMXBs. In the accompanying paper, Ivanova et al. (2012) present theoretical interpretations of our finding, supported by numerical simulations.

## 2. SAMPLE

We extracted our primary sample of early type galaxies from the ACS Virgo cluster survey (ACSVCS; Cote et al. 2004) and the ACS Fornax cluster survey (ACSFCS; Jordan et al. 2007). Both data sets provide lists of GCs, which were homogeneously analyzed for their photometry in two optical bands with F475W (g) and F850LP (z) and half-light radii in each band (Jordan et al. 2009, Virgo; Jordan 2012 in prep., Fornax). Because the ACS field of view is 202" x 202", the entire galaxy (inside the $D_{25}$ ellipse) is included only for a few small galaxies. For most galaxies, the outer regions are excluded (see Table 1). We discuss the effect of sampling different spatial region on our results in section 6. For NGC 4649, we use the GC data from Strader et al. (2012) who have analyzed five additional ACS pointings to cover the entire galaxy. Their photometry and sizes are consistent with those of Jordan et al. (2009) for the GCs in common.

Table 1. Sample Early Type Galaxies

| Galaxy | type | d (Mpc) | R25 (') | $M_B$ (mag) | log $L_K$ ($L_K$ sun) | Reference |
|---|---|---|---|---|---|---|
| **Main Sample** | | | | | | |
| N4365 | E3 | 23.1 | 3.4 x 2.5 | -21.33 | 11.40 | ACSVCS |
| N4374 | E1 | 18.5 | 3.2 x 2.8 | -21.33 | 11.38 | ACSVCS |
| N4472 | E2 | 16.7 | 5.1 x 4.1 | -21.78 | 11.62 | ACSVCS |
| N4621 | E4 | 14.9 | 2.6 x 1.8 | -20.34 | 10.98 | ACSVCS |
| N4649 | E2 | 16.5 | 3.7 x 3.0 | -21.39 | 11.47 | Strader12 |
| N1399 | E0 | 20.9 | 3.4 x 3.2 | -21.16 | 11.45 | ACSFCS |
| N1404 | E2 | 20.2 | 1.6 x 1.4 | -20.55 | 11.21 | ACSFCS |
| N1427 | E4 | 19.6 | 1.8 x 1.2 | -19.65 | 10.66 | ACSFCS |
| **Additional sample for faint LMXBs** | | | | | | |
| N3379 | E1 | 10.6 | 2.7 x 2.4 | -19.94 | 10.87 | Kim09 |
| N4278 | E1 | 16.1 | 2.0 x 1.9 | -20.06 | 10.87 | Kim09 |
| N4649 | E6 | 11.8 | 3.6 x 2.3 | -20.28 | 10.93 | Kim09 |
| Cen A | S0 pec | 4.2 | 12.9 x10.0 | -20.82 | 11.00 | Voss09 |
| M31 | SAb | 0.76 | 99.3 x30.8 | -21.04 | 10.70 | Voss09 |
| **Additional sample for brightest LMXBs** | | | | | | |
| M87[a] | E0 | 16.7 | 4.1 x 3.3 | -21.62 | 11.45 | ACSVCS |
| **Additional sample for ULXs** | | | | | | |
| N1316[b] | S0 pec | 21.0 | 6.0 x 4.2 | -22.21 | 11.74 | ACSFCS |
| N1380[b] | S0/a | 21.2 | 2.3 x 1.1 | -20.71 | 11.24 | ACSFCS |

reference
ACSVCS = ACS Virgo Cluster Survey; ACSFCS = ACS Fornax Cluster Surgey; Strader12 = Strader et al. (2012) for N4649; Kim09 = Kim et al. (2009) for 3 elliptical galaxies with deep Chandra observations; Voss09 = Voss et al. (2009) for Cen A and the M31 bulge.

a. Because of the complex gas structure in M87, we only use bright LMXBs with $L_x > 5 \times 10^{38}$ erg s$^{-1}$ (in 0.3-8keV) or net count > 300 after visual inspection of individual sources.
b. These galaxies are used only for ULXs ($L_x > 2 \times 10^{39}$ in 0.3-8keV) in Table 6.

We obtained the X-ray data from the public *Chandra* Archive (http://asc.harvard.edu/cda). Starting with the ten brightest, GC-richest elliptical galaxies in each cluster, we selected our sample galaxies with more than ten GC-LMXBs: NGC 4365, 4374, 4472, 4621, and 4649 from the Virgo cluster and NGC 1399, 1404 and 1427 from the Fornax cluster. We also used our proprietary *Chandra* data (PI: Fabbiano) taken in 2011 of NGC 4649 (see Luo et al 2012). We excluded NGC 1316 (S0 pec) and 1380 (S0/a), which could add 9 and 13 GC-LMXBs, respectively, because they may contain a younger stellar population. In our homogeneously old sample, the optical color can be assumed to be a reasonable indicator of metallicity (e.g., Peng et al. 2006; Mieske et al. 2010). In Table 1, we list the basic galaxy information, including morphological types from RC3, distances from Blakeslee et al. (2009), semi-major and semi-minor axes of the D25 ellipse, B magnitudes from RC3, and K-band luminosities from 2MASS.

Additionally, we used the published results for GC-LMXBs in NGC 3379, 4278 and 4697 from Kim et al. (2009), and in Cen A (NGC 5128) and the bulge of M31 from Voss et al. (2009) to supplement the number of very faint LMXBs (section 5.2). We used M87 (NGC 4486, Virgo A, the dominant galaxy of the Virgo cluster) to increase the number of very luminous LMXBs ($L_X > 5 \times 10^{38}$ erg s$^{-1}$; section 5.3). Although M87 hosts a large number of GCs, the X-ray point sources can be confused with false detections at the low luminosities, because of the complex spatial distribution of the hot gas. However, very luminous LMXBs (with high S/N) are exempt from this problem.

Table 2. Chandra Observations

| Galaxy | ObsId | Observation Dates | Exposure (ksec) | $L_X^a$ ($10^{37}$ erg/s) |
|---|---|---|---|---|
| N4365 | 2015, 5921-4, 7224 | Jun 2001 – Nov 2005 | 191 | 2.6 |
| N4374 | 803, 5908, 6131 | May 2000 – Nov 2005 | 110 | 5.0 |
| N4472 | 321, 11274, 12888-9 | Jun 2000 – Feb 2011 | 359 | 3.0 |
| N4621 | 2068 | Aug 2001 | 23 | 6.5 |
| N4649 | 785, 8182, 8507, 12975-6, 14328 | Apr 2000 – Aug 2011 | 280 | 2.6 |
| N1399 | 319, 9530 | Jan 2000 – Jan 2008 | 114 | 6.8 |
| N1404 | 2942, 9798, 9799 | Feb 2003 – Dec 2007 | 65 | 13.1 |
| N1427 | 4742 | May 2005 | 50 | 7.2 |
| N1316$^b$ | 2022 | Apr 2001 | 24 | 1.4 |
| N1380$^b$ | 9526 | Mar 2008 | 38 | 6.5 |

a. limiting X-ray luminosity (in 0.3-8keV) at 50% detection probability
b. used only for ULXs (see Table 6)

## 3. *CHANDRA* X-RAY DATA ANALYSIS

For all the galaxies, we only used data obtained with the S3 chip of *Chandra* Advanced CCD Imaging Spectrometer (ACIS; Garmire 1997), in order to take advantage of the higher response of the back-illuminated CCD. We excluded short (< 10 ksec) ACIS-I exposures of NGC 1399 and NGC 4472, to avoid mixing data from different detectors. In all cases, the ACIS temperature was −120 C at which the ACIS calibration is most reliable. The basic information of the *Chandra* observations is summarized in Table 2, which includes observation ID, observation dates and net exposure times. Many galaxies were observed multiple times between 2000 and 2011, with individual exposures ranging from 20 to 100 ks. Also listed in Table 2 is the limiting luminosity for X-ray point source detection at 50% probability. Since we are measuring the ratio of GC-LMXB fractions between red and blue GCs, there is no need to correct for the incompleteness of LMXBs detection (see section 6).

The ACIS data were uniformly reduced with a custom-made pipeline (XPIPE), specifically developed for the Chandra Multi-wavelength Project (ChaMP; Kim et al. 2004). Starting with the CXC pipeline level 2 products, we apply *acis_process_events* available in CIAO v4 with up-to-date calibration data, to correct for the time and position-dependent gain and QE variation. After removing background flares, we re-project individual observations to a common tangent point and combine them using merge_all from the CIAO contributed package. We use wavdetect to detect X-ray point sources, setting the exposure threshold to be 10% using an exposure map, to eliminate the false detections often found at the chip edge; we also set the significance threshold to be $10^{-6}$, corresponding approximately to one false source per chip. The performance and limitations of wavdetect are well understood and calibrated by extensive simulations (e.g., Kim & Fabbiano 2003; Kim et al. 2004; Kim, M. et al. 2007). To measure the broad-band 0.3–8 keV X-ray luminosity, we calculate the energy conversion factor (ECF = ratio of flux to count rate) for each source in each observation, by assuming a power-law spectral model with a photon index of $\Gamma = 1.7$ (e.g., Irwin et al. 2003; Boroson et al. 2011) and Galactic $N_H$ (Dickey & Lockman 1990). For sources detected in the merged data, we apply an exposure-weighted mean ECF. For X-ray luminosities of LMXBs, we use the broad energy band in 0.3–8 keV throughout this paper.

## 4. MATCHING LMXBs and GCs

To obtain our sample of LMXBs in GCs, we identified GC-LMXBs matches by cross-correlating the X-ray and optical source lists. We first determined the systematic positional offset between the samples of X-ray and optical sources. After correcting for the systematic shift (up to ~1"), we applied a strict matching criterion with a radius of 0.5". This procedure is simpler than that used in Kim et al. (2009), where field LMXBs and background sources were also identified. Here, we are only concerned with GC-LMXBs, hence we select only clean GC-LMXB matches. Although more GC-LMXBs can be identified with a relaxed search radius, the probability of a spurious identification increases significantly. The chance probability of a false match within 0.5" is 0.5-2% (≤ 1 false match per galaxy). The number of duplicated matches also significantly increases with a larger search radius. With our search radius (0.5"), we have 7 duplicated matches, where a single X-ray source is matched with two GCs. We use these sources in our sample only if both matching GCs belong to the same (blue or red) color subsample.

Based on the bimodal GC color distribution, the division in red and blue GCs occurs at g-z = 1.1-1.2 (see Peng et al, 2006 for Virgo galaxies and Mieske et al. 2010 for Fornax galaxies). Because of the host galaxy dependency, the average color of GCs (both blue and red GCs) increases (becomes redder) with increasing $M_B$ (or color) of the host galaxy. Our sample consists mostly of giant early type galaxies with $M_B < -20.5$ mag or $L_K > 10^{11}$ $L_{Ksun}$ (excluding the smallest galaxy N1427; see Table 1) and the division between blue and red GCs occurs in a narrow range of g-z. Therefore, we select the division at g-z=1.15 for all galaxies in our sample of ACSVCS and ACSFCS. We also tried slightly different division colors (g-z fixed at 1.1 or 1.2, or variable within 1.1 and 1.2 depending on $M_B$), but our results (the ratio of GC-LMXB fractions in red and blue GCs) do not vary in any significant manner.

```
            Table 3 Number of GCs and GC-LMXBs
            ---------------------------------------
                         GC              GC-LMXB
            Galaxy    ALL R-GC B-GC    ALL R-GC B-GC
            ---------------------------------------
            N4365     907  505  402     68   52   16
            N4374     506  221  285     17   10    7
            N4472     765  478  287     70   65    5
            N4621     308  171  137     16   14    2
            N4649    1603  890  713    141  119   22

            N1399    1074  709  365     63   51   12
            N1404     380  217  163     11   11    0
            N1427     361  137  224     22    9   13

            sum      5904 3328 2576    408  331   77

            GC-LMXB fraction^a        0.069 0.099 0.030
            error                     0.003 0.006 0.003
            ---------------------------------------

            a. fraction of GCs hosting LMXBs
```

In Table 3, we list the number of GCs and GC-LMXBs for each of the eight Virgo and Fornax elliptical galaxies. Out of 5904 GCs, 408 (or 6.9%) host LMXBs. This GC-LMXB fraction is a lower limit, because some of the *Chandra* observations used are quite shallow (see the limiting $L_X$ in Table 2 and section 6 for more discussions). The total number of red GCs is slightly higher than that of the blue GCs in our samples. We note that the ratio may change depending on the sampling regions and the color boundary between red and blue GCs. However, the ratio of GC-LMXB fractions is not seriously affected by these variations; see section 6 for more discussions.

## 5. COMPARISON BETWEEN LMXB IN RED AND BLUE GCs

To determine GC-LMXB fractions as a function of $L_X$, we count the numbers of LMXBs in red and blue GCs in seven luminosity bins with the minimum $L_X = 10^{36}$ erg s$^{-1}$, and calculate the ratio of GC-LMXB fractions in red and blue GCs in each bin:

$$R = \frac{N(RGC-LMXB)/N(RGC)}{N(BGC-LMXB)/N(BGC)}$$

In Table 4, we summarize the results. We plot the color-magnitude diagram in Figure 1, where red and blue GCs with LMXBs are marked by red and blue colors, respectively. We also plot the histogram of GCs as a function of g-z in Figure 2, which clearly indicates the higher GC-LMXB fraction in red GCs.

### 5.1. INTERMEDIATE LUMINOSITY LMXBs

The ratio, $R$, is best determined in the intermediate $L_X$ ranging from 2 x $10^{37}$ to 5 x $10^{38}$ erg s$^{-1}$ (in the middle four $L_X$ bins). $R$ ranges from 2.7 to 4.5. To quantitatively measure the significance of *R being larger than unity*, we calculate the error associated with $R$ using the Bayesian estimators, BEHR (Park et al. 2006) which rigorously treats the Poisson statistics taking into account the non-Gaussian nature of the error. We list the lower and upper bounds at the 68% confidence level and the confidence level at which $R > 1$ in Table 4. In the middle four $L_X$ bins ($L_X$ = 2 x $10^{37}$ - 5 x $10^{38}$ erg s$^{-1}$), the significance of $R$ being larger than unity is $3\sigma$ (99.7%) or higher. If we count all GC-LMXBs in this $L_X$ range, the ratio and its allowed range is well determined to be:

$<R> = 3.4 \pm 0.5$ for GC-LMXBs with $L_X$ = 2 x $10^{37}$ - 5 x $10^{38}$ erg s$^{-1}$.

$R$ in these middle four $L_X$ bins stays consistent with the above $<R>$ within a $1\sigma$ deviation (see Figure 3), clearly indicating that the metallicity effect on the GC-LMXB fraction is *independent of $L_X$ of LMXBs* in this $L_X$ range within our uncertainties.

```
Table 4 GC-LMXB fractions in red and blue GCs (from Virgo and Fornax galaxies)
-----------------------------------------------------------------------------
                   N(GC-LMXB) in a given L_X bin (in 10^37 erg/s)^a
       N(GC)   0.1-1  1-2   2-5    5-10   10-20  20-50  50-     2-50   all
-----------------------------------------------------------------------------
red    3328    4      8     60     86     80     75     18      301    331
blue   2576    0      2     17     18     21     13     6       69     77

R              -      3.1   2.7    3.7    2.9    4.5    2.3     3.4    3.3
Lower (68%)    0      0.6   1.9    2.7    2.2    3.1    1.1     2.9    2.9
Upper (68%)    23.9   5.3   3.4    4.6    3.6    5.8    3.3     3.8    3.7
Conf (%)^b                  99.91  99.99+ 99.99+ 99.99+ 72
-----------------------------------------------------------------------------
a. L_X in 0.3-8keV
b. The confidence level at which R > 1.
```

### 5.2. LOW LUMINOSITY LMXBs

In the first two bins ($L_X < 2 \times 10^{37}$ erg s$^{-1}$), the face value of the ratio is consistent with the above value, but $R$ is poorly constrained due to the small number of faint sources. To improve the statistics at the low $L_X$ end, we used published results of the very deep observations of NGC 3379, 4278 and 4649 (see the luminosity functions in Kim et al. 2009), which have resulted in the faintest detections of LMXBs in this type of galaxies (down to a few x $10^{36}$ erg s$^{-1}$). We also use the results of Cen A and the M31 bulge from Voss et al. (2009). Although Cen A and M31 are not normal elliptical galaxies, the stellar populations of their bulges are similarly old; these observations provide even fainter LMXBs (down to $10^{36}$ erg s$^{-1}$) because of their proximity. Our results with data from these additional five galaxies are summarized in Table 5. While the results are almost identical to those in Table 4 in high $L_X$ bins, the statistics are significantly improved in the first two bins. The ratios *R* are 2.1 and 2.7 in the first two bins, and the 1σ allowed range is 1.2 – 3.7. The significance of *R* being larger than unity is now 85% and 94% in the first 2 bins. If we combine the two bins, we get *R* = 2.5 and the 1σ range is 1.6 – 2.9. The significance of *R* being larger than unity is 99%, which we consider as marginally significant.

```
Table 5 same as table 4, but with additional galaxies[a]
-----------------------------------------------------------------------
                 N(GC-LMXB) in a given L_X bin (in 10^37 erg/s)[b]
       N(GC)   0.1-1  1-2   2-5   5-10  10-20  20-50   >50[c]   2-50   all
-----------------------------------------------------------------------
red    4127    26     27    84    110   97     82     30        373    446
blue   3294    10      8    27     24   28     18      9         97    121

R              2.1    2.7   2.5   3.7   2.8    3.6    2.5        3.1    2.9
Lower (68%)    1.2    1.5   1.9   2.8   2.2    2.7    1.4        2.7    2.6
Upper (68%)    2.7    3.7   3.0   4.4   3.4    4.6    3.4        3.4    3.2
Conf (%)[d]    85     94    99.99 99.99+ 99.99+ 99.99+ 95
-----------------------------------------------------------------------

a. Data for N3379, N4278, N4649, M31, and Cen A are from the literature
b. L_X in 0.3-8keV
c. The last bin (L_X > 5 x 10^38) also includes GC-LMXBs from M87.
d. The confidence level at which R > 1.
```

### 5.3. HIGH LUMINOSITY LMXBs

In the high $L_X$ bin ($L_X > 5 \times 10^{38}$ erg s$^{-1}$) in Table 4, the ratio *R* is 2.5, but with a large error. The significance of *R* being larger than unity is only ~1σ (72%). Using data from NGC 3379, 4278 and 4649 (Kim et al. 2009) and from Cen A and the bulge of M31 (Voss et al. 2009), does not help because there are only two luminous LMXBs in these galaxies. To further improve statistics, we add GC-LMXBs in M87 (NGC 4486) with extra cautions. At lower $L_X$, we avoided LMXBs in M87 because of spatial incompleteness and possible false detections resulting from the complex nature of its strong hot gas X-ray emission (e.g., Million et al. 2010). However, for most luminous LMXBs ($L_X > 5 \times 10^{38}$ erg s$^{-1}$, corresponding to ~300 net counts), false identifications are unlikely. Analyzing two long ACIS-S Chandra observations of M87 (obsid=2707 and 3717, with total exposure of 130 ksec) with the techniques described in section 4, we identified 17 candidate GC-LMXBs with $L_X > 5 \times 10^{38}$ erg s$^{-1}$. After visual inspection of the images, we eliminated three LMXBs in red GCs that may be small-scale gas clumps rather than point sources and an LMXB that matches both a blue and a red GC within 0.5". We obtained additional 10 (3) LMXBs in 1078 (667) red (blue) GCs. Adding them to those in Table 4, we have 30 (9) LMXBs in 5205 (3961) red (blue) GCs. The ratio, $R$ is 2.5 and the 1$\sigma$ allowed range is 1.4 – 3.4. The significance of $R$ being larger than unity is at 95% (or 2$\sigma$).

This trend may continue to the highest luminosity range, i.e., to the Ultra Luminous X-ray sources (ULX) whose X-ray luminosity, $L_X > 2 \times 10^{39}$ erg s$^{-1}$, is significantly higher than the Eddington luminosity of stellar mass black holes (e.g., Fabbiano 2006). In our sample, we have 7 ULXs among GC-LMXBs with $L_X > 2 \times 10^{39}$ erg s$^{-1}$; 5 (2) of them are found in red (blue) GCs (see Table 6). While the ratio between red and blue GC-ULXs (ULX residing in GC) is $R \sim 3$, the sample is too small to draw any significant conclusion. For completeness, we add to Table 6 a few possible ULXs in GCs that are previously known to have the peak luminosity at least $L_X = 2 \times 10^{39}$ erg s$^{-1}$ (even if the average $L_X$ is lower): a red GC-ULX in NGC 1316 (Kim & Fabbiano 2003), a blue GC-ULX in NGC 3379 (Brasssington et al. 2012) and two (in red and blue GC each) in NGC 4472 (Maccarone et al, 2007, 2011). The two sources in N4472 were not included in section 4, because they are outside the ACS fov. We also found one ULX in a red GC in NGC 1380 that we excluded in the above analysis (see section 2). Combining all known GC-ULXs, we have 8 and 4 ULXs residing in red and blue GCs, respectively.

Table 6 ULX in GCs

| name | z mag | g-z mag | d kpc | $L_X$ $10^{38}$ | ref |
|---|---|---|---|---|---|
| **ULX in red GC** | | | | | |
| N1399 | 21.28 | 1.37 | 6.67 | 24.3 | |
| N1399 | 20.49 | 1.60 | 4.15 | 30.4 | |
| N4486 | 22.55 | 1.61 | 4.72 | 21.7 | |
| N4486 | 20.29 | 1.34 | 0.51 | 23.4[a] | |
| N4649 | 20.26 | 1.55 | 8.97 | 23.8 | |
| **ULX in blue GC** | | | | | |
| N1427 | 21.30 | 1.03 | 5.35 | 21.2 | |
| N4486 | 21.68 | 1.08 | 0.56 | 25.3[a] | |
| **Additional GC-ULXs from the literature** | | | | | |
| N1316[c] | 22.13 | 1.60 | 1.29 | 34[a] | 1 |
| N1380[c] | 22.30 | 1.37 | 5.55 | 35.6 | this work |
| N3379 | 21.9 | 0.8 | 10.14 | 28[b] | 2 |
| N4472 | - | blue | 32.56 | 45[b] | 3 |
| N4472 | - | red | 13.08 | 20[b] | 4 |

a   very close to the galaxy center and the $L_X$ could be affected by the strong diffuse emission.
b   peak luminosity
c   some GCs may not be old

reference 1. Kim & Fabbiano (2003); 2. Brassington et al, 2012); 3. Maccarone et al. (2007); 4. Maccarone et al. (2011)

If we count all GC-LMXBs ($L_X > 10^{36}$ erg s$^{-1}$), the ratio and its allowed range is (the last column in Table 5):

$$<R> = 2.9 \pm 0.3 \quad \text{for GC-LMXBs with } L_X > 10^{36} \text{ erg s}^{-1}.$$

In summary (see Figure 3), a metallictiy dependent effect (i.e., $R > 1$) is confirmed in four independent $L_X$ bins at the intermediate luminosity $L_X = 2 \times 10^{37} - 5 \times 10^{38}$ erg s$^{-1}$. In this $L_X$ range, $R$ stays constant and consistent within 1σ errors with the weighted average $<R> = 3.4 \pm 0.5$ (also with $<R> = 2.9 \pm 0.3$ that was determined from the entire $L_X$ range), indicating that the metallicity effect on the GC-LMXB fraction is *independent of $L_X$ of LMXBs* in this $L_X$ range. At the fainter $L_X$ ($< 2 \times 10^{37}$) and brighter $L_X$ ($> 5 \times 10^{38}$), the significance of the metallicity dependence is marginal (at 2σ); while the ratio is still consistent with $R \sim 3$, the uncertainty is larger. To estimate the possible variation of $R$ as a function of $L_X$, we fit the data in a functional form of $R = a \log(L_X) + b$. The best fit value and error of the slope is $a = -0.1 \pm 0.7$ for $L_X > 10^{37}$ erg s$^{-1}$ in the main sample (table 4 or blue filled circles in Fig 3) and $a = 0.3 \pm 0.4$ for $L_X > 10^{36}$ erg s$^{-1}$ in the extended sample (table 5 or black open circles in Fig 3). These results reinforce our claim that the current data show no clear dependence of $R$ on $L_X$; however, we note that some dependence is still allowed given the error on slope. Our results suggest that the XLFs of LMXBs in red and blue GCs have a similar shape, but that their normalizations differ by a factor of ~3. The available data at the lowest $L_X$ ($10^{36}$ - $10^{37}$ erg s$^{-1}$) and highest $L_X$ ($> 2 \times 10^{39}$ erg s$^{-1}$) are consistent with this conclusion; however, more LMXBs are needed to statistically confirm this.

## 6. DISCUSSION

Because the optical color of stellar systems is primarily determined by two parameters, age and metallicity, it is important to select a homogeneous age sample so that separating GCs by color yields different metallicity samples. As described in section 2, our main sample consists of pure, old elliptical galaxies in the Virgo and Fornax clusters. The average stellar age of our sample galaxies ranges from 8 Gyr to 14 Gyr (Trager et al. 2000, Terlevich & Forbes 2002, Thomas et al. 2005, Sanchez-Blaquez et al. 2006, Idiart et al. 2007, Serra & Oosterloo 2010). We note that the age of an old elliptical galaxy is more reliable and less subject to systematic error (stemming from the common assumption of single stellar population) than that of a disturbed rejuvenated galaxy where most stars are old and only a small fraction of younger stellar population may be mixed in (e.g., Serra & Oosterloo 2010). Since our sample is not expected to be affected by age differences as a function of metallicity, we can apply the color metallicity, (g-z)-[Fe/H], relation presented in Peng et al. (2006) to translate the color difference to metallicity difference[1]: the color difference, Δ(g-z) = 0.4-0.5 between the two peaks of blue and red GCs (Figure 2) corresponds to a mean metallicity difference of a factor of 10-30.

A few studies on the spectroscopic ages of a limited number of GCs also indicate that the majority of GCs in our sample galaxies have old ages of ≳10 Gyr (Forbes et al. 2001 for NGC 1399, Beasley et al. 2000 for N4472, Pierce et al. 2006 for NGC 4649). However, there has been some debate on the possible presence of intermediate age GCs in NGC 4365 (Brodie et al. 2005; Kundu et al. 2005), while the average stellar age is old (12.6 Gyr, Sanchez-Blaquez 2006; 11.3 Gyr, Idiart et al. 2007). The recent near-IR study by Chies-Santos et al. (2011) suggests that there is no significant population of young GCs in NGC 4365. Nonetheless, after excluding NGC 4365, we confirm that our results remain valid. Another possible case of contamination that might arise from a younger stellar population is Cen A. It could affect $R$ in the lowest $L_X$ bin. However, the spectroscopic study by Beasley et al. (2008) showed that the majority of their sample (~90%) is old and only a small fraction may have intermediate ages (4-6 Gyr). We also note that the ratio of GC-LMXB fractions in Cen A is similar to that of the bulge of M31 in the lowest $L_X$ bin.

In addition to the metallicity effect on the GC-LMXB fraction, it is well known that massive GCs are more likely to host LMXBs (e.g., Sarazin et al. 2003; Kim, E. et al. 2006). This is driven by their large number of stars and dense cores, which result in higher dynamical encounter rates (Jordan et al. 2007). In our sample, this effect is shown in Figure 1. In particular, about half of the GCs with $M_Z < -11.5$ mag host LMXBs. We also note that all of these most massive GCs with LMXBs are red. We therefore need to examine if the metallicity effect may be driven by the mass of GCs. In Figure 4, we show the histogram of the absolute magnitude ($M_Z$) of red and blue GCs (solid lines for all GCs). If we assume a Gaussian distribution, the mean $<M_Z>$ = -8.82 mag and -8.66 mag for red and blue GCs, respectively, and σ = 1.1 for both samples. The median values are also similar (-8.73 and -8.57 mag). The red GCs may be systematically more massive than the blue GCs, but the difference is only 0.16 mag in mean (also in median) values of the two samples, which corresponds to ~15% difference in luminosity or mass. This is clearly too small to affect our results in the different GC-LMXB fraction (by a factor of ~3) in red and blue GCs.

Another important parameter is the stellar encounter rate, Γ, which takes both mass and compactness into account; the higher the encounter rate, the higher the probability to dynamically form LMXBs in GCs (e.g., Sivakoff et al, 2007; Jordan et al. 2007; Peacock et al. 2010). We tested whether red GCs have systematically higher Γ. In that case, Γ could be the driving force. Following Sivakoff et al. (2007), the stellar encounter rate can be written

$$\Gamma \sim M_{GC}^{1.5} r_h^{-2.5},$$

---

[1] We note that the precise form of the relation between [Fe/H] and (g-z) color is not known precisely. Although the precise form of the relation can have a large impact on the interpretation of [Fe/H] distributions derived from color (see, e.g., Blakeslee et al 2012) the relation presented in Peng et al (2006) is adequate for the purposes of this work.

where $M_{GC}$ and $r_h$ are the GC mass and the half-mass radius. We calculate the stellar encounter rates for all GCs and compare them in red and blue GC samples (see Figure 5.) The encounter rates in red and blue GC samples are statistically identical; their mean values differ only by 4%, much smaller than σ (~0.8 in log Γ), assuming a Gaussian distribution. This clearly indicates that while the stellar encounter rate is important in forming LMXBs, it is completely independent of the metallicity effect on the GC-LMXB fraction. As expected, on average, Γ is higher by a factor of several in GCs with LMXBs than GCs without LMXBs. And this trend is similar in both red and blue GCs. To further illustrate the last point, we re-select blue and red GCs with the same encounter rate (Γ = $5 \times 10^6 - 5 \times 10^7$). In the $L_X$ range of $2 \times 10^{37} - 5 \times 10^{38}$ erg s$^{-1}$, we recover $R = 3.2 \pm 0.6$, which is fully consistent with the result in section 5.

Because we sample GCs in different regions of different galaxies, we also considered whether there is any spatial factor that may affect our results. Given that we consider the ratio of GC-LMXB fractions in red and blue GCs, any systematic problem is not important as long as it affects the red and blue GCs in the same manner. One possible issue may be a combined effect caused by the different radial distribution of red and blue GCs and position-dependent completeness of LMXBs. It is well known that the distribution of blue GCs is more spatially extended (with a flatter radial profile) than that of red GCs (e.g., Strader et al. 2012). For example, the new ASC data of NGC 4649 by Strader et al. (2012), which cover the entire galaxy, results in 130% more blue GC and 80% more red GCs, than the ACSVCS data (202" x 202") of Jordan et al. (2007). As the exposure time of each X-ray observation varies from one galaxy to another, the limiting luminosity in detecting LMXBs varies (Table 2). The completeness in detecting point sources primarily depends on the background emission and the PSF, which work in an opposite sense as a function of galacto-centric distance. While it is harder to detect X-ray sources near the center of the galaxy where the gas emission is stronger and point sources are more crowded, the PSF becomes wider with increasing off-axis angle, also affecting the threshold sensitivity. Another related issue is the position dependent GC size (e.g., Masters et al. 2010) such that GCs are more compact near the center of the galaxy. To quantitatively assess the combined effect on $R$ and to confirm our conclusions, we re-measured the ratio of GC-LMXB fractions, $R$, separately in galactocentric regions with radius r < 60" and r = 60 – 120". We find that $R$ is consistent between two regions; $R = 2.9 \pm 0.6$ and $3.5 \pm 0.7$ in the inner and outer regions, respectively (or $3.3 \pm 0.7$ and $3.5 \pm 0.8$ when only LMXBs with $L_X = 2 \times 10^{37} - 5 \times 10^{38}$ are used). In addition, we measured $R$ with GC-LMXBs of N4649 outside the central ACSVCS fov (r = 2' – 4.5'), and again we find that $R$ (= 3.2) in the outer regions. This assures that our results of a fairly constant ratio of the GC-LMXB fraction in red and blue GCs is robust.

Because different types of LMXBs (with different compact sources and with different donors) dominate at different luminosities (e.g., Fragos et al. 2008), our result of a metallicity effect that holds across a wide range of $L_X$ strongly implies a mechanism that must be effective for multiple types of LMXBs. In particular, magnetic breaking may work for faint ($L_X < 2 \times 10^{37}$ erg s$^{-1}$) LMXBs with main-sequence donors (Ivanova 2006), but cannot work for luminous LMXBs, which are likely to have red giant donors or WD donors. Therefore our result requires a new explanation, which is discussed in a companion paper by Ivanova et al. (2012).

## 7. SUMMARY

We measured the ratio of GC-LMXB fractions in red and blue GCs as a function of X-ray luminosity, to assess if there is a luminosity-dependent metallicity effect on the presence of LMXBs in GCs. We find that the ratio between the fraction of LMXBs in red and blue GCs, $R$, is close to 3 in a wide range of $L_X$, indicating that the metallicity effect is independent of $L_X$. The statistical significance of metallicity dependence (i.e., $R > 1$) is solid in the intermediate luminosity $L_X = 2 \times 10^{37} - 5 \times 10^{38}$ erg s$^{-1}$. At the fainter $L_X$ ($10^{36} - 2 \times 10^{37}$) and brighter $L_X$ ($> 5 \times 10^{38}$), the significance of the metallicity dependence is marginal (at 2$\sigma$); however the ratio that is still consistent with $R \sim 3$. This trend appears to continue to the ULX regime, but with a large uncertainty. Our results suggest that the XLFs of LMXBs in red and blue GCs have a similar shape. We note that the metallicity effect is not affected by the stellar (e.g., age), GC dynamical (e.g., stellar encounter rate) properties and/or the observational selection effect (e.g., different sample regions). Our result requires a new explanation for the $L_X$-independent metallicity effect on the dynamical formation of LMXBs in GCs.

The data analysis was supported by the CXC CIAO software and CALDB. We have used the NASA NED and ADS facilities, and have extracted archival data from the Chandra archives. This work was supported by NASA contract NAS8-39073 (CXC) and Chandra Guest Observer grant GO1-12110X (PI Fabbiano). NI and GRS acknowledge support by NSERC Discovery Grants, with additional support for NI by the CRC Program. TF acknowledges support from the CfA and the ITC prize fellowship programs. AJ acknowledges support from the Chilean Ministry for the Economy, Development, and Tourism's Programa Iniciativa Científica Milenio through grant P07-021-F, awarded to The Milky Way Millennium Nucleus, from BASAL CATA PFB-06 and from Anillo ACT-086. This material is based upon work supported in part by the National Science Foundation Grant No. 1066293 and the hospitality of the Aspen Center for Physics.

# REFERENCES

Beasley, M. A., Sharples, R. M., Bridges, T. J., Hanes, D. A., Zepf, S. E., Ashman, K. M., & Geisler, D. 2000, MNRSA, *318*, 1249

Beasley, M. A., Bridges, T., Peng, E., Harris, W. E., Harris, G. L. H., Forbes, D. A., & Mackie, G. 2008, MNRAS, *386*, 1443

Bellazzini, M., Pasquali, A., Federici, L., Ferraro, F. R., & Pecci, F. F. 1995, ApJ, 439, 687

Bildsten, L., & Deloye, C. J. 2004, ApJ, 607, L119

Blakeslee, J. P., Jordan, A., Mei, S., Cote, P., Ferrarese, L., Infante, L., Peng, E. W., et al. 2009, ApJ, *694*, 556

Blakeslee, J. P. et al. 2012, ApJ, 746, 88

Boroson, B., Kim, D.-W., & Fabbiano, G. 2011, ApJ, 729, 12

Brassington, N. J., Fabbiano, G., Zezas, A., Kundu, A., Kim, D.-W., Fragos, T., King, A. R., et al. 2012, ApJ, in press

Brodie, J. P., Strader, J., Denicoló, G., Beasley, M. A., Cenarro, A. J., Larsen, S. S., Kuntschner, H., et al. 2005, ApJ, 129, 2643

Brodie J. P., Strader J., 2006, ARA&A, 44, 193

Chies-Santos A. L., Larsen S. S., Kuntschner H., Anders P., Wehner E. M., Strader J., Brodie J. P., Santos J. F. C., 2011, A&A, 525, A20

Cote, P. et al. Cote, P., Blakeslee, J. P., Ferrarese, L., Jordan, A., Mei, S., Merritt, D., Milosavljević, M., et al. 2004, ApJS, 153, 223

Dickey, J. M. and Lockman, F. J. 1990, ARAA, 28, 215

Fabbiano, G. 2006 ARAA, 44, 323

Fragos, T., Kalogera, V., Belczynski, K., Fabbiano, G., Kim, D. âW., Brassington, N. J., Angelini, L., et al. 2008, ApJ, 683, 346

Forbes, D. A., Beasley, M. A., Brodie, J. P., & Kissler-Patig, M. (2001). ApJ *563*, L143

Garmire, G. P. 1997, AAS, 190, 3404

Grindlay, J. E. 1984, Adv. Space Rev., 3, 19

Idiart, T. P., Silk, J., & de Freitas Pacheco, J. A. 2007, MNRAS, 381, 1711

Irwin, J. 2005, ApJ, 631, 511

Irwin, J., et al. 2003, ApJ, 587, 356

Irwin, J., et al. 2004, ApJ, 601, L143

Ivanova, N. 2006, ApJ, *636*, 979

Ivanova, N., Heinke, C. O., Rasio, F. A., Belczynski, K., & Fregeau, J. M. 2008, MNRAS, 386, 553

Jordan, A., Blakeslee, J. P., Cote, P., Ferrarese, L., Infante, L., Mei, S., Merritt, D., et al. 2007, ApJS, *169*, 213

Jordan, A., Sivakoff, G. R., McLaughlin, D. E., Blakeslee, J. P., Evans, D. A., Kraft, R. P., Hardcastle, M. J., et al. 2007, ApJL, *671*, L117

Jordan, A., Peng, E. W., Blakeslee, J. P., Cote, P., Eyheramendy, S., Ferrarese, L., Mei, S., et al. 2009, ApJS, *180*, 54

Juett, A. M. 2005, ApJ, 621, L25

Kim, D.-W., & Fabbiano, G. 2003, ApJ, 586, 826

Kim, D.-W., et al. 2004, ApJS, 150, 19

Kim, D.-W., Fabbiano, G., Brassington, N. J., Fragos, T., Kalogera, V., Zezas, A., Jordan, A., et al. (2009).ApJ, *703*, 829

Kim, E., Kim, D.-W., Fabbiano, G., Lee, M. G., Park, H. S., Geisler, D., & Dirsch, B. 2006, ApJ, *647*, 276
Kim, M., et al. 2007, ApJS, 169, 401
Kundu, A., Maccarone, T. J., & Zepf, S. E. 2002, ApJ, 574, L5
Kundu, A., Zepf, S. E., Hempel, M., Morton, D., Ashman, K. M., Maccarone, T. J., Kissler-Patig, M., et al. 2005, ApJ, 634, L41
Kundu, A., Zepf, S. E., Hempel, M., Morton, D., Ashman, K. M., Maccarone, T. J., Kissler-Patig, M., et al. 2005, ApJ, 634, L41
Luo, B., Fabbiano, G., Strader, J., Kim, D., Brodie, J. P., Fragos, T., Gallagher, J. S., et al. 2012, ApJS, submitted
Maccarone, T. J., Kundu, A., & Zepf, S. E. 2004, ApJ, *606*, 430
Maccarone, T. J., Kundu, A., Zepf, S. E., & Rhode K. L., 2007, Nat, 445, 183
Maccarone, T. J., Kundu, A., & Zepf, S. 2011, MNRAS, 410, 1655
Masters, K. L., Jordán, A., Côté, P., Ferrarese, L., Blakeslee, J. P., Infante, L., Peng, E. W., et al. 2010, ApJ, 715, 1419
Mieske, S., Jordan, A., Cote, P., Peng, E. W., Ferrarese, L., Blakeslee, J. P., Mei, S., et al. 2010, ApJ, *710*, 1672
Million, E. T., Werner, N., Simionescu, A., Allen, S. W., Nulsen, P. E. J., Fabian, A. C., Böhringer, H., et al. (2010, MNRAS, *407*, 2046
Paolillo, M., Puzia, T. H., Goudfrooij, P., Zepf, S. E., Maccarone, T. J., Kundu, A., Fabbiano, G., et al. 2011, ApJ, 736, 90.
Park, T., Kashyap, V. L., Siemiginowska, A., van Dyk, D. A., Zezas, A., Heinke, C., & Wargelin, B. J. 2006, ApJ, *652*, 610
Peacock, M. B., Maccarone, T. J., Kundu, A., & Zepf, S. E. 2010, MNRAS, 407, 2611
Peng, E. W., Jordan, A., Cote, P., Blakeslee, J. P., Ferrarese, L., Mei, S., West, M. J., et al. 2006, ApJ, *639*, 95
Pierce, M., Bridges, T., Forbes, D. A., Proctor, R., Beasley, M. A., Gebhardt, K., Faifer, F. R., et al. 2006, MNRAS, *368*, 325
Revnivtsev, M., Postnov, K., Kuranov, A., & Ritter, H. 2011, A&A, 526, A94
Sanchez-Blazquez, P., Gorgas, J., Cardiel, N., & Gonzalez, J. J. 2006, A&A 457, 787
Sarazin, C. L., Kundu, A., Irwin, J. A., Sivakoff, G. R., Blanton, E. L., & Randall, S. W. 2003, ApJ, 595, 743
Serra, P., & Oosterloo, T. A. 2010, MNRAS, 401, L29
Sivakoff, G. R., Jordan, A., Sarazin, C. L., Blakeslee, J. P., Cote, P., Ferrarese, L., Juett, A. M., et al. 2007, ApJ, *660*, 1246
Strader, J., Fabbiano, G., Luo, B., Kim, D.-W., Brodie, J. P., Fragos, T., Gallagher, J. S., et al. 2012, ApJ, submitted
Terlevich, A. I., & Forbes, D. A 2002, MNRAS, 330, 547
Thomas, D., Maraston, C., Bender, R., & Mendes de Oliveira, C. 2005, ApJ, 621, 673
Tonry, J. L., Dressler, A., Blakeslee, J. P., et al. 2001, ApJ, 546, 681
Trager, S. C., Dressler, A., Blakeslee, J. P., et al. 2000, AJ, 120, 165
Verbunt, F., & Lewin, W. H. G. 2006, in Compact Stellar X-ray Sources, ed. Lewin & van der Klis (Cambridge: Cambridge Univ. Press), 341
Verbunt, F., & van den Heuvel, E. P. J. 1995, in X-ray Binaries, ed. W. H. G. Lewin, J. van Paradijs, & E. P. J. van den Heuvel (Cambridge: Cambridge Univ. Press), 457

Voss, R., Gilfanov, M., Sivakoff, G. R., Kraft, R. P., Jordán, A., Raychaudhury, S., Birkinshaw, M., et al. (2009).ApJ, *70*, 471

Zhang, Z., Gilfanov, M., Voss, R., Sivakoff, G. R., Kraft, R. P., Brassington, N. J., Kundu, A., et al. 2011, A&A, *533*, 33

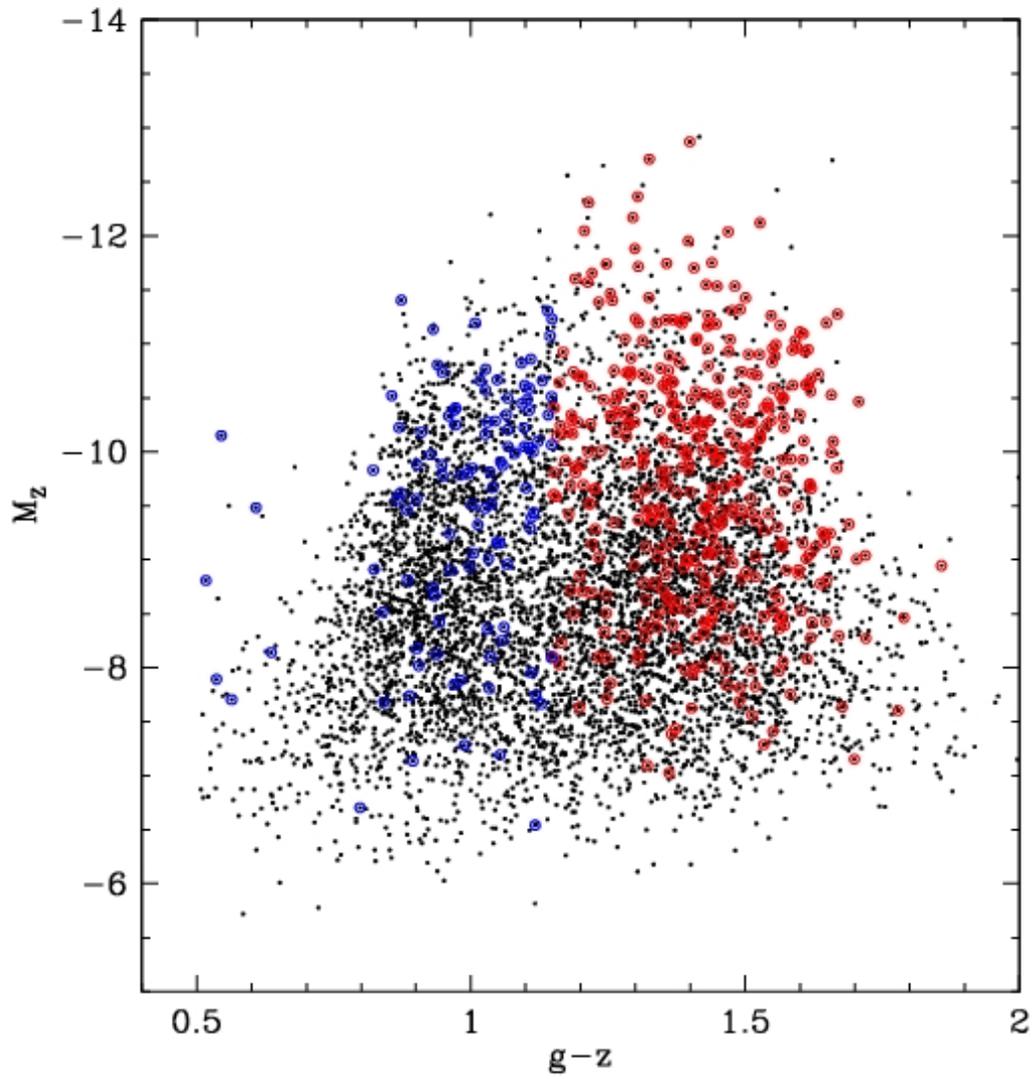

Figure 1. Color-magnitude diagram of GCs. The red and blue GCs with LMXBs are separately highlighted by red and blue circles.

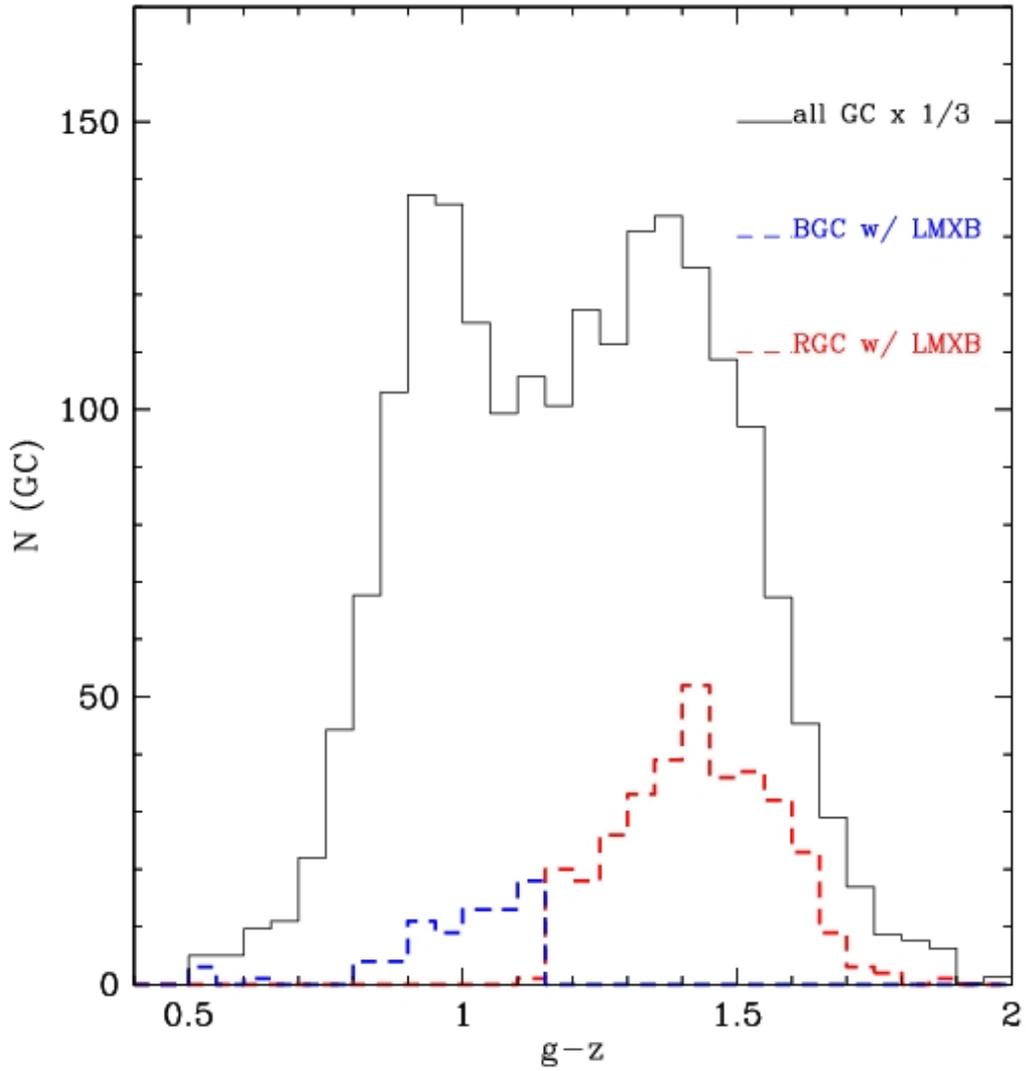

Figure 2. Histogram of GCs as a function of g-z color. The black line indicates for all GCs, but scaled down by a factor of 1/3 for clarity. The red and blue GCs with LMXBs are separately plotted by red and blue lines. The GC-LMXB fraction strongly depends on the GC color.

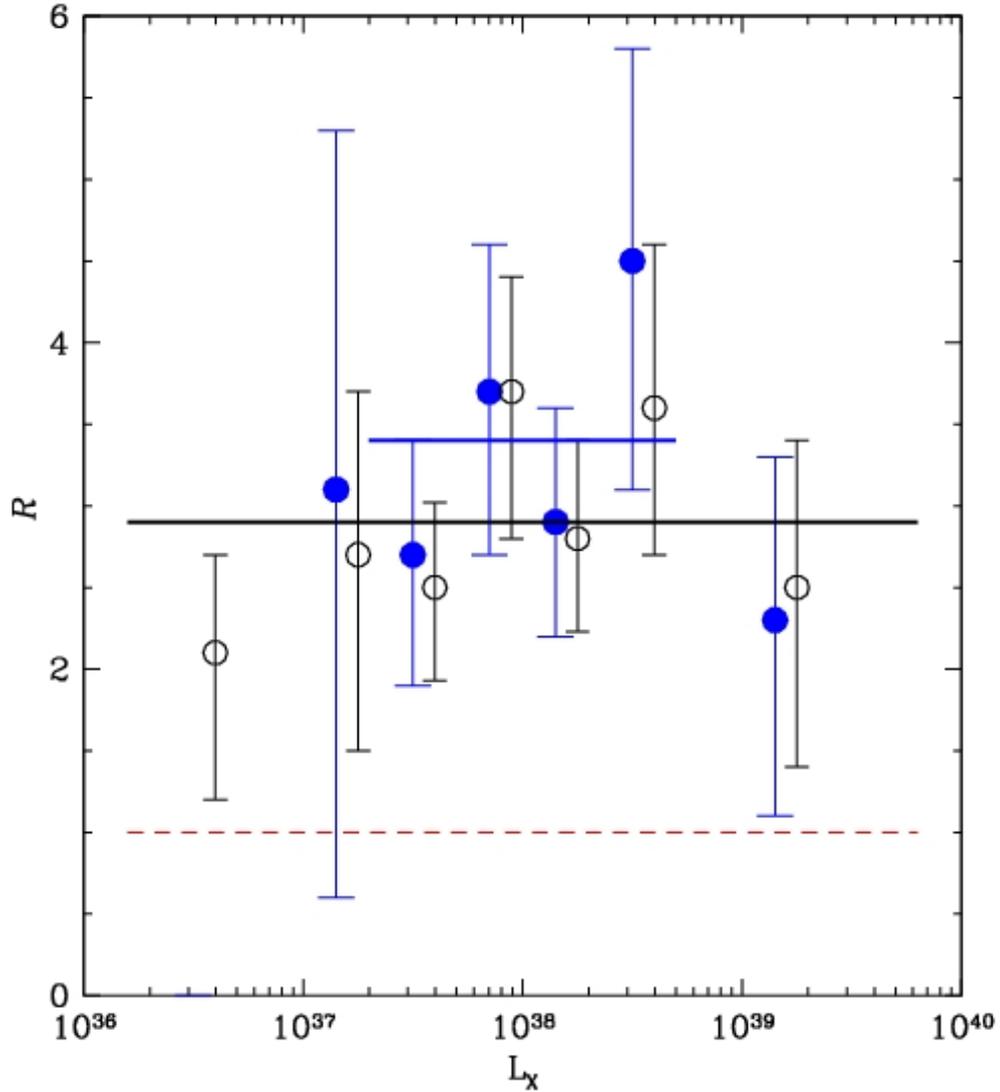

Figure 3. The ratio of GC-LMXB fractions between red and blue GCs, *R*, as a function of $L_X$ of LMXBs. The blue circles were determined from the main sample of the Virgo and Fornax galaxies (Table 4) and the black circles (slightly shifted horizontally for clarity) from the extended sample (Table 5). <*R*> in $L_X = 2 \times 10^{37} - 5 \times 10^{38}$ erg s$^{-1}$ from the main sample (and in $L_X > 10^{36}$ erg s$^{-1}$ from the extended sample) is marked by the blue (and black) horizontal line. The red dashed line indicate the condition when the GC-LMXB fraction is the same in red and blue GCs.

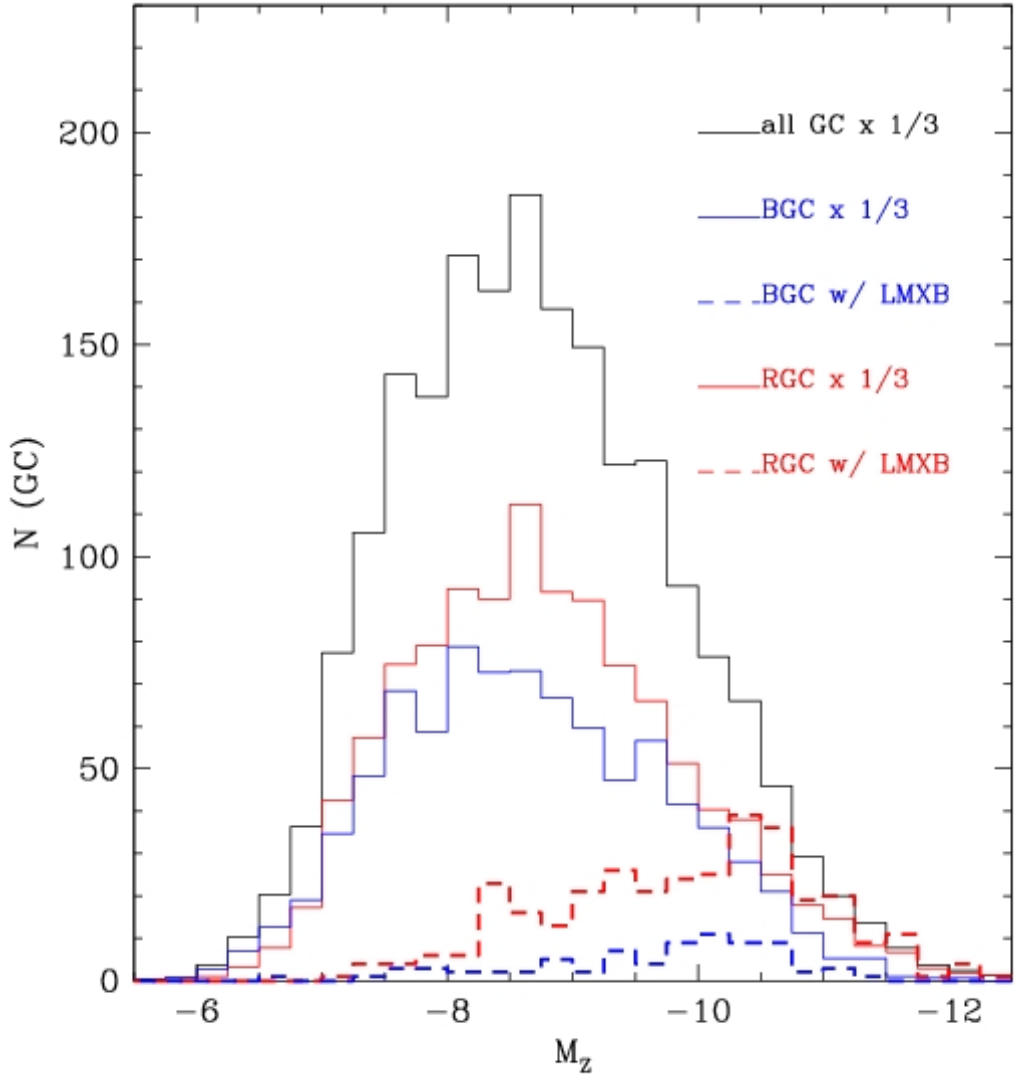

Figure 4. Histogram of GCs as a function of absolute magnitude $M_Z$. The red and blue GC samples are separately plotted by red and blue lines. The solid lines are for all GCs (but scaled down by a factor of 1/3 for clarity) and the dashed lines for only GCs with LMXBs.

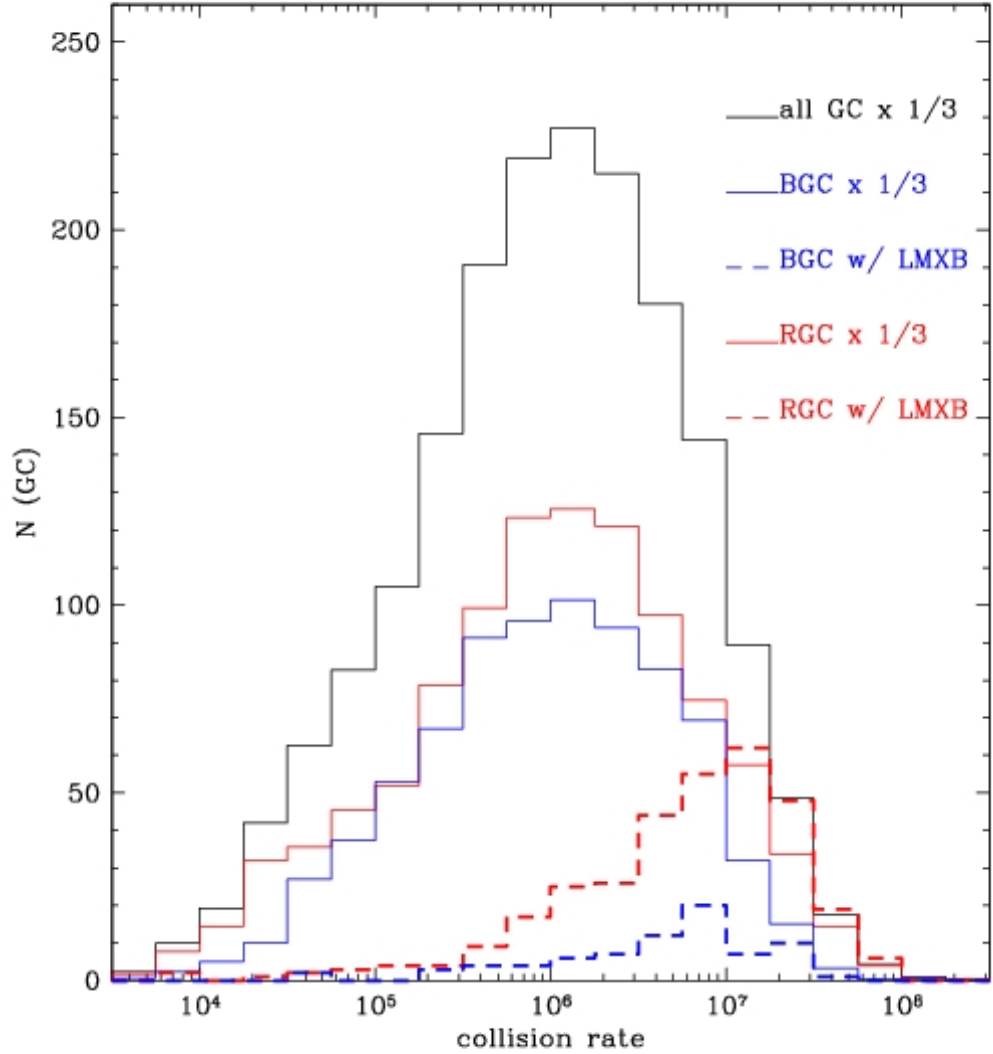

Figure 5. Histogram of GCs as a function of stellar encounter rate ($\Gamma$). The red and blue GC samples are separately plotted by red and blue lines. The solid lines are for all GCs (but scaled down by a factor of 1/3 for clarity) and the dashed lines for only GCs with LMXBs.